\documentclass[twocolumn,showpacs,preprintnumbers,amsmath,amssymb,aps]{revtex4}

\usepackage{graphicx}
\usepackage{dcolumn}
\usepackage{bm}

\begin{document}

\preprint{}

\title{Scalar phantom energy as a cosmological dynamical system}

\author{L. Arturo Ure\~na-L\'opez}
\email{lurena@fisica.ugto.mx}
\affiliation{%
Instituto de F\'{\i}sica de la Universidad de Guanajuato,\\
 A.P. E-143, C.P. 37150, Le\'on, Guanajuato, M\'exico
}%


\date{\today}

\begin{abstract}
Phantom energy can be visualized as a scalar field with a
(non-canonical) negative kinetic energy term. We use the dynamical
system formalism to study the attractor behavior of a cosmological
model containing a phantom scalar field $\phi$ endowed with an
exponential potential of the form $V(\phi)=V_0 \exp(-\lambda \kappa
\phi)$, and a  perfect fluid with constant equation of state $\gamma$;
the latter can be of the phantom type too. As in the canonical case,
three characteristic solutions can be identified. The scaling solution
exists but is either unstable or of no physical interest. Thus, there
are only two stable critical points which are of physical interest,
corresponding to the perfect fluid and scalar field dominated
solutions, respectively. The most interesting case arises for $0>
\gamma > -\lambda^2/3$, which allows the coexistence of the three
solutions. The main features of each solution are discussed in turn.  
\end{abstract}

\maketitle

\section{Introduction}
\label{sec1}
For many years, scalar fields have played the role of work-horses for
many phenomenological models in modern Cosmology. And the main issue
is to find which solutions of the field equations are physically
important to explain the universe we live in.

The usual route was to get expertise in solving the coupled
Einstein-Klein-Gordon (EKG) set of nonlinear differential equations, as it
can be seen in the early literature on scalar fields. However, it was
recently realized that for a exponential potential the EKG equations
could be seen as a plane autonomous system of
equations\cite{Copeland:1997et}. That is, the EKG equations could be
seen as a dynamical system. This case became one more among the many
examples of dynamical systems in Cosmology, see\cite{dynasystems}.

Even though the exponential potential is one of the most studied cases
in the cosmology of scalar fields
(see\cite{Sahni:2004ai,Ohta:2004wk,Neupane:2003cs,Chen:2003ij} and references
therein), the way it is handled in\cite{Copeland:1997et} allows a
better understanding of the evolution of a universe with exponential
potentials. The idea has been exploited in some papers that have used
it successfully for diverse situations. For instance, in the cases of negative
exponentials\cite{Heard:2002dr}, an exponential potential in general
Robertson-Walker spacetimes\cite{vandenHoogen:1999qq}, and
multi-exponential potentials\cite{Guo:2003eu,Collinucci:2004iw}; see
also\cite{Vieira:2003tx}.

On the other hand, the idea of a scalar field with non-canonical
kinetic energy has drawn the attention of cosmologists because such a
scalar field can have, in principle, what is called a \textit{phantom}
equation of
state\cite{Caldwell:1999ew,Gibbons:2003yj,Cline:2003gs,Carroll:2003st,Nojiri:2005sr,Neupane:2005nb},
for which $p/\rho < -1$, being $p$ and $\rho$ the pressure and the
energy density of the phantom fluid, respectively. This new type of
fluids violates the so-called dominant energy condition, which for a
perfect fluid is written as $|p| \leq \rho$; for a thoroughly
discussion on this condition and the stability of vacuum
see\cite{Cline:2003gs,Carroll:2003st}. Thus, the case of phantom scalar
fields should be taken with care, as some even think that a
cosmological constant may be easier to
justify\cite{Gibbons:2003yj,Cline:2003gs,Carroll:2003st}. The presence
of a kind of phantom matter may be supported by supernovae type Ia
observations, as phantom matter seems to fit them better than a
cosmological constant or a quintessence
field~\cite{Jassal:2005qc,Alam:2004jy,Jassal:2004ej}. Other
alternatives, though, are a canonical scalar field climbing up its
scalar potential\cite{Csaki:2005vq,Sahlen:2005zw}, quantum
corrections on large scales\cite{Onemli:2004mb,Onemli:2002hr}, and a
non-canonical complex scalar field\cite{Wei:2005nw};
mechanisms that may result in a phantom-like equation of state.

Recently, the attractor behavior of phantom scalar field models, which
have a negative kinetic term, was studied in\cite{Guo:2004ae} using
the Hamilton-Jacobi formalism. Such study was restricted to
cosmological models in which the phantom scalar field was the only
matter present, using three different types of scalar potentials
(power-law, exponential and cosine).

In the present paper, our aim is to perform a more complete study of the
attractor properties of a phantom scalar field endowed with a positive
exponential potential, using the dynamical system formalism developed
in\cite{Copeland:1997et} for canonical scalar fields. This study will
include a perfect fluid with a constant barotropic equation of
state.

In this respect, it should be said that a similar analysis was
performed in\cite{Hao:2003ww,Gumjudpai:2005ry}, where a coupling between the
phantom scalar field and a dust fluid was included. The main
difference with that work is that we will not consider any coupling,
and will not restrict the perfect fluid equation of state to the
usual values; we will allow the possibility of super-stiff and phantom
perfect fluids.

There are many other authors that have studied the cosmologies
containing a phantom field, and have obtained explicit
solutions together with careful stability analysis,
see\cite{Guo:2004ae,Hao:2003ww,Gumjudpai:2005ry,Li:2005ay,Elizalde:2004mq,Nojiri:2005sx,Piazza:2004df,Singh:2003vx,Hao:2003th,Li:2003ft}
and the long list of references there in. We will comment on any
similarity or difference that may appear with respect to those papers.

A brief summary of the contents of this paper is as follows. In
Sec~\ref{sec2}, we will review the mathematics needed for a
cosmological model with a perfect fluid and a phantom scalar field. In
Sec.~\ref{sec3}, the critical points of the system and their stability
are presented. The discussion is focused in the three different cases
that appear according to the values the perfect-fluid equation of
state can take. Finally, Sec.~\ref{sec4} is devoted to conclusions.

\section{Mathematical background}
\label{sec2}
For simplicity, we shall restrict ourselves to the case of a
homogeneous and isotropic universe with zero curvature, its metric
given by the flat Robertson-Walker one\footnote{We will be using a
  metric signature $(-,+,+,+)$, and units in which $c=1$}. In this
universe, there is a perfect fluid with pressure $p_\gamma$ and energy
density $\rho_\gamma$ related through a constant equation of state
$\gamma$ in the form $p_\gamma = (\gamma -1)\rho_\gamma$. There is
also a phantom scalar field $\phi$ minimally coupled to gravity, whose
Lagrangian density reads $\mathcal{L} = (1/2) \partial^\mu \phi
\partial_\mu \phi - V(\phi)$, where $V(\phi)$ is called the scalar
potential. Therefore, the equations of motion are the known EKG
equations
\begin{subequations}
\label{basic}
\begin{eqnarray}
\dot{H} &=& -\frac{\kappa^2}{2} \left( \gamma \rho_\gamma -
\dot{\phi}^2 \right) \, , \\
\ddot{\phi} &=& -3H\dot{\phi} + \frac{dV}{d\phi} \, , \\
\dot{\rho_\gamma} &=& -3H\gamma \rho_\gamma \, ,
\end{eqnarray}
\end{subequations}
together with the constraint (Friedmann) equation
\begin{equation}
H^2 \equiv \left( \frac{\dot{a}}{a} \right)^2 = \frac{\kappa^2}{3}
\left( \rho_\gamma - \frac{1}{2} \dot{\phi}^2 +V \right) \, ,
\label{fried}
\end{equation}
where dots mean derivative with respect to cosmic time, $\kappa^2=8\pi
G$, $H$ is the Hubble parameter, and $a(t)$ is the scale factor of the
universe. The above equations are the same as obtained in the
canonical case except for the minus sign in front of the scalar
kinetic term $\dot{\phi}^2$. 

We will study the case of a scalar exponential potential of the
form $V(\phi)=V_0 \exp(-\lambda \kappa \phi)$, where $V_0$ and
$\lambda$ are free parameters. The similarity with the canonical case
suggests that we should make use of the dimensionless
variables\cite{Copeland:1997et}
\begin{equation}
x\equiv \frac{\kappa \dot{\phi}}{\sqrt{6}H} \, , \quad y \equiv
\frac{\kappa \sqrt{V}}{\sqrt{3}H} \, , \label{newvars}
\end{equation}
and then Eqs.~(\ref{basic}) can be written as a plane-autonomous system
\begin{subequations}
\label{autoeqs}
\begin{eqnarray}
x^\prime &=& -3x + \frac{3}{2} \left[ \gamma \left( 1+x^2-y^2
 \right) -2x^2 \right] x - \sqrt{\frac{3}{2}} \lambda y^2 \, ,
 \label{autoeqsa} \\
y^\prime &=&  \frac{3}{2} \left[ \gamma \left( 1+x^2-y^2
 \right) -2x^2 \right] y - \sqrt{\frac{3}{2}} \lambda x y \, ,
 \label{autoeqsb}
\end{eqnarray}
\end{subequations}
where now a prime denotes derivative with respect to the logarithm of the
scale factor $N\equiv \ln a$. The constraint equation~(\ref{fried})
now reads
\begin{equation}
1= \frac{\kappa^2 \rho_\gamma}{3H^2} - x^2 +y^2 \, . \label{fried1}
\end{equation}

There are two scalar quantities we shall be interested in, which are
the phantom density parameter $\Omega_\phi$ and the effective
phantom equation of state $\gamma_\phi$, given by
\begin{equation}
\Omega_\phi = -x^2+y^2 \,, \quad \Omega_\phi \gamma_\phi = -2 x^2 \, .
\end{equation}
We notice that the scalar equation of state $\gamma_\phi$ is always negative.

It is interesting to note that there will be three different regions
in the phase plane, characterized by $\Omega_\phi < 0$, $0 \leq
\Omega_\phi \leq 1$, and $\Omega_\phi > 1$, respectively. The boundary
lines between the above regions are given by the hyperbola $y^2-x^2=1$
and the straight lines $y=\pm x$, and are shown with dashed curves in
the figures below.

As we will focus our attention in $\rho_\gamma >0$, we see the
Friedmann constraint~(\ref{fried}) requires all realistic trajectories
in the phase plane $\{x,y\}$ to comply with $\Omega_\phi \leq
1$. However, in opposition to the canonical case, there is no reason
to constraint $\Omega_\phi$ only to positive values. As we shall see
later, the existence of perfect-fluid \emph{supra-dominated} eras cannot
be discarded.

\section{Critical points and stability}
\label{sec3}
Eqs.~(\ref{autoeqs}) are written as an autonomous
phase system of the form $\mathbf{x}^\prime = \mathbf{f}(\mathbf{x})$,
and its so-called critical points $\mathbf{x}_0$ are solutions of the
system of equations $\mathbf{f}(\mathbf{x}_0)=0$. To determine their
stability we need to perform linear perturbations around the critical
points in the form $\mathbf{x}=\mathbf{x}_0+\mathbf{u}$, which results
in the following equations of motion $\mathbf{u}^\prime = \mathcal{M}
\mathbf{u}$, where
\begin{equation}
\mathcal{M}_{ij} = \left. \frac{\partial f_i}{\partial x_j}
\right|_{\mathbf{x}_0} \, . \label{eigenmatrix}
\end{equation}
A non-trivial critical point is called stable (unstable) whenever the
eigenvalues $m_i$ of $\mathcal{M}$ are such that $\mathrm{Re}(m_i) <0$
($\mathrm{Re}(m_i) > 0$). If neither of the aforementioned cases is
accomplished, the critical point is called a saddle
point\cite{dynasystems}. 

We applied the above procedure to Eqs.~(\ref{autoeqs}). The critical
points together with the stability analysis, and the corresponding
values of $\Omega_\phi$ and $\gamma_\phi$ are shown in
Table~\ref{t:critical}\footnote{There are other two critical points
  $\mathbf{x}_0=(\pm i,0)$, which are the counterparts of the kinetic
  dominated solutions in the canonical case. But, they are obviously
  not of physical interest this time, see also Table~I
  in\cite{Gumjudpai:2005ry}.}. For completeness, the eigenvalues of
matrix~(\ref{eigenmatrix}) for the critical points A, B, C and D are
shown in Table~\ref{t:eigen1}. For solution E, the eigenvalues are
\begin{equation}
m_\pm = -\frac{3}{4} \left| 2-\gamma \right| \left[
  \frac{(2-\gamma)}{\left| 2-\gamma \right|} \pm \sqrt{1-\frac{8\gamma
  \left( \lambda^2+3\gamma \right)}{\lambda^2\left( 2-\gamma \right)}}
  \right] \, . \label{pointE}
\end{equation}

\begin{table*}[t]
\caption{ \label{t:critical} Critical points for a phantom scalar
  field endowed with an exponential potential.}
\begin{ruledtabular}
\begin{tabular}{cccp{1.4cm}p{3.7cm}p{0.6cm}c}
Label & $\rho_\gamma$ & $\mathbf{x}_0$ & \centering Existence &
\centering Stability & \centering $\Omega_\phi$ & $\gamma_\phi$ \\ \hline 
A & $\rho_\Lambda$ & $(0,0)$ & \centering $\forall \lambda$ &
\centering Saddle & \centering $0$ & Undefined \\
B & $\rho_\Lambda$ & $\left( -\lambda/\sqrt{6},\sqrt{1+\lambda^2/6} \right)$ &
\centering $\forall \lambda$ & \centering Stable & \centering $1$ &
$-\lambda^2/3$ \\ \hline
C & $>0$ & $(0,0)$ & \centering $\forall \lambda$, $\forall \gamma$ &
\centering Unstable for $\gamma > 2$ \newline Saddle for $2 > \gamma > 0$ 
\newline Stable for $\gamma < 0$ & \centering $0$ & Undefined \\
D & $>0$ & $\left( -\lambda/\sqrt{6},\sqrt{1+\lambda^2/6} \right)$ &
\centering $\forall \lambda$, $\forall \gamma$ & \centering Stable for
$\gamma > -\lambda^2/3$ \newline Saddle for $\gamma < -\lambda^2 /3$ &
\centering $1$ & $-\lambda^2/3$ \\
E & $>0$ & $\left( \sqrt{3/2}\gamma/\lambda,\sqrt{3\gamma
  (\gamma-2)/2\lambda^2} \right)$ & $\forall \lambda$, $\gamma > 2$ \newline
$\forall \lambda$, $\gamma < 0$ & \centering Saddle for $\gamma >2$
\newline Saddle for $0 > \gamma > -\lambda^2/3$ \newline Stable for
$\gamma < -\lambda^2 /3$ & \centering $< 0$ \newline $-3\gamma
/\lambda^2$ \newline $> 1$ & $\gamma$
\end{tabular}
\end{ruledtabular}
\end{table*}

\begin{table}
\caption{ \label{t:eigen1} Eigenvalues found for the critical points
  A, B, C and D shown in Table~\ref{t:critical}, see Eq.~(\ref{eigenmatrix}).}
\begin{ruledtabular}
\begin{tabular}{ccc}
Label & $m_{-}$ & $m_{+}$ \\ \hline 
A & $-3$ & $0$ \\
B & $-3-\lambda^2/2$ & $-\lambda^2$ \\ \hline
C & $-3+3\gamma /2$ & $3\gamma/2$ \\
D & $-3-\lambda^2/2$ & $-3\gamma-\lambda^2$
\end{tabular}
\end{ruledtabular}
\end{table}

There are many similarities in the critical points with respect to the
canonical case, so we briefly review the main solutions in the latter,
see Table~I in\cite{Copeland:1997et}. There are three solutions that
are relevant for Cosmology, which are respectively dubbed the perfect
fluid dominated solution, the scalar field dominated solution, and the
scaling solution (for an interesting discussion on the existence of
scaling solutions, see\cite{Piazza:2004df}).

The fluid dominated solution is in general unstable, except in the
case of a cosmological constant. The scalar field dominated solution
only exists for $\lambda^2 < 6$, in which case is the only late-time
attractor and is inflationary for $\lambda^2 < 2$. Also, this solution
is a saddle point when a cosmological constant is present. 

For both cases $3\gamma < \lambda^2$ and $6 < \lambda^2$, the only
stable solution is the scaling one, in which the scalar energy density
remains proportional to that of the perfect fluid with $\Omega_\phi =
3\gamma /\lambda^2$. This last property has been widely used in some
models of scalar field dark
matter\cite{Matos:2004rs,Sahni:1999qe}\footnote{It should be noticed
  that the \emph{scaling} solution E in Table~\ref{t:critical} is very
  similar to the so called \emph{tracker} attractor solution for phantom
  fields\cite{Hao:2003th}; though the latter does not exist for an
  exponential potential.}.

As we shall see, the minus sign of the kinetic energy introduces
relevant changes for the phantom case. There are three different
situations we are about to revise now.

\subsection{$\gamma = 0$}
As it was found in\cite{Hao:2003ww,Gumjudpai:2005ry}, in a universe with a
cosmological constant $\rho_\gamma = \rho_\Lambda$ and a phantom
scalar field there is only one stable critical point, which is the
scalar field dominated solution B. This the same as in the canonical
case, except that now there is not restriction on the values that
$\lambda$ can take\cite{Hao:2003ww,Gumjudpai:2005ry}. As we mentioned
before, this also implies that an exponential potential will come to
dominate over a cosmological constant too; this is to be expected
since $\gamma_\phi < 0$.

\subsubsection{Only phantom scalar field}
It should be realized that the case in which no other matter except
the phantom scalar field is present is \emph{not} equivalent to the above
case with $\gamma =0$. In the former, Eqs.~(\ref{autoeqs}),
and~(\ref{fried1}) can be fused into the single equation
\begin{equation}
x^\prime = - \left( 3x+\sqrt{\frac{3}{2}} \lambda \right) \left( 1+x^2
\right) \, .
\end{equation}

Clearly, the only physically reasonable critical point is the same as
  point B\cite{Elizalde:2004mq}\footnote{This is at variance with the
  results in\cite{Guo:2004ae}, where the authors claim that there is
  indeed the same restriction as in the canonical case, namely
  $\lambda^2 < 6$, for the existence of this solution. In our study,
  we have found no evidence for such restriction, see also
  Fig.~\ref{fig:caseD} below.}, which is again stable. This time,
  however, point A cannot exist.

\subsection{$\gamma > 0$}
\begin{figure}
\includegraphics[width=8cm]{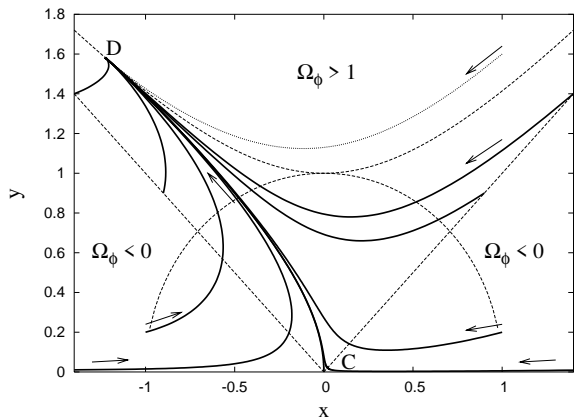}
\caption{\label{fig:caseD} The phase plane for $\gamma=1$ and
  $\lambda=3$. For $\gamma \geq 0$, the only late-time attractor is the stable
  critical point D. Notice that, in opposition to the canonical scalar
  field, there is not restriction on the value of $\lambda$. We also
  show the phantom-divide (dashed) circumference, and some trajectories where
  the perfect fluid is \emph{supra-dominant}, see text for details. For
  completeness, we show a representative trajectory with $\Omega_\phi
  >1$ (dotted curves), which never enters the physically interesting
  region $\Omega_\phi \leq 1$. For the case in which there is
  not a perfect fluid present, all trajectories lie on the hyperbola
  $y^2-x^2=1$ and finish at the critical point D. }
\end{figure}

On the other hand, when a ($\gamma \neq 0$) perfect fluid is present,
things become more interesting. For $\gamma >0$, the scaling solution
E either does not exist or is not of physical interest, and the
perfect fluid dominated solution C is in general unstable. Therefore,
we conclude that the (phantom) scalar field dominated solutions B and
D are unavoidable if $\gamma \geq 0$, which is the case for ordinary
matter (radiation, dust, and even quintessence fields and a
cosmological constant)\cite{Hao:2003ww}. The effective equation of
state $\gamma_\phi$ is of the phantom type, and the scalar field
climbs up its scalar potential. This is shown in Fig.~\ref{fig:caseD}
for $\gamma=1$.

This is the first example in this work where we see
\emph{supra-dominance} of the perfect fluid over the phantom field,
i.e. $\Omega_\gamma > 1$. Such a phenomenon, that passed unnoticed in
previous published works\cite{Elizalde:2004mq,Hao:2003ww}, also arises
if the perfect fluid and the phantom field are coupled, as
in\cite{Gumjudpai:2005ry}.

Also in Fig.~\ref{fig:caseD}, we plotted trajectories that start with
$\Omega_\phi =0$ (at the boundary lines $y=\pm x$). We believe that
all of such trajectories are indeed part of longer trajectories which
may have started in the negative-$\Omega_\phi$ regions.

It is also worth noticing that trajectories can start with $\Omega_\phi <
0$, but always end up with $0\leq \Omega_\phi \leq 1$, but not
vice-versa; this is because there are not stable critical points with
$\Omega_\phi < 0$ (in the case of coupling between matter and the
phantom field, there are stable critical points in which the perfect
fluid matter is supra-dominant, see\cite{Gumjudpai:2005ry}). Moreover,
the parabola $y^2-x^2=1$ is not traversable at all, as trajectories
with $\Omega_\phi > 1$ never enters into the region $0\leq \Omega_\phi
\leq 1$ in order to reach the stable critical point C.

To better visualize  the evolution of the universe along the
trajectories in Fig.~\ref{fig:caseD}, it is convenient to define an
effective equation of state of the universe $\gamma_\textrm{eff}$,
which for our case reads
\begin{equation}
\gamma_\textrm{eff} \equiv \frac{\gamma \rho_\gamma + \gamma_\phi
  \rho_\phi}{\rho_\gamma + \rho_\phi} = \gamma \left[ 1 - y^2 -
  \left(\frac{2-\gamma}{\gamma} \right) x^2 \right] \, . \label{geff}
\end{equation}
The value $\gamma_\textrm{eff}=0$ defines, in general, an ellipse
(hyperbola) in the $xy$ plane if $0 < \gamma < 2$ ($\gamma > 2$). That
is to say, this ellipse (and this hyperbola) marks the phantom divide
of the universe in the phase plane: $\gamma_\textrm{eff} >0$
($\gamma_\textrm{eff} < 0$) inside (outside) the ellipse (and the
hyperbola).

For $\gamma =1$, the phantom divide is a unitary circumference, as shown in
Fig.~\ref{fig:caseD}. A curious point is that some trajectories cross
over the phantom divide twice. Even though the perfect fluid can be
supra-dominant, the effective equation of the universe rarely coincides
with that of the perfect fluid.

\subsection{$\gamma < 0$}
\begin{figure}
\includegraphics[width=8cm]{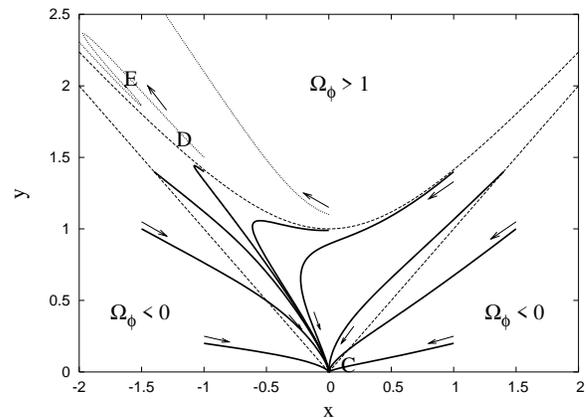}
\caption{\label{fig:caseC} The phase plane for $\gamma=-4$ and
  $\lambda=3$. For this case, the only late-time attractor is the stable
  critical point C, i.e., the (phantom) perfect fluid dominates over the
  scalar field, even if a stage of supra-dominance is allowed
  initially. For completeness, we show numerical solutions of the case
  $\Omega_\phi > 1$ (dotted curves), for which point E is an attractor stable
  solution, which is in agreement with the stability analysis in
  Eq.~(\ref{pointE}).}
\end{figure}

There is an important change in the analysis above if we allow for
$\gamma < 0$, that is, if the perfect fluid is also phantom. For $\gamma < -
\lambda^2 /3$, D is a saddle point (the scalar field dominated
solution cannot be sustained), and point E is again of no physical
interest even if it is a stable node. The only possibility is then the
(phantom) perfect fluid dominated solution C (the perfect fluid is more
phantom than the scalar field). The phase plane for an instance of
this case is shown in Fig.~\ref{fig:caseC}.

\begin{figure}
\includegraphics[width=8cm]{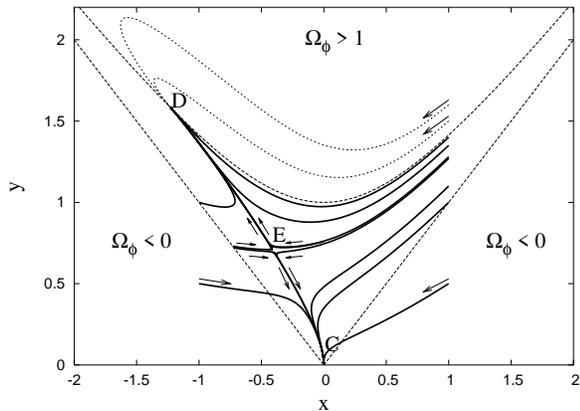}
\caption{\label{fig:caseCD} The phase plane for $\gamma=-1$ and
  $\lambda=3$. For this case, there are \emph{two} late-time attractors,
  which are the critical points C and D; though the saddle point E is
  also present. Which of the stable critical points is reached at the
  end, depends on the initial values $\mathbf{x}_0$. Notice that point
  D is also a late-time attractor for $\Omega_\phi > 1$ (dotted curves).}
\end{figure}

There is still one extra case we have to care about. For $0> \gamma >
-\lambda^2/3$, both points C and D exist and are
\textit{stable}. Moreover, point E is also physically allowed, but this
time is a saddle point. The numerical experience, as shown in
Fig.~\ref{fig:caseCD}, indicates that some of the trajectories pass
nearby point E before finishing at one of the late-attractors C and
D. How close a trajectory is to point E depends on the initial
conditions $\mathbf{x}_0$. 

We also see that all trajectories with initial perfect fluid
supra-dominance never reach point D, and then always end up at critical
point C. As for the $\Omega > 1$ case, we see that all trajectories
end up at point D, but again without crossing over the $\Omega_\phi \leq 1$ region.

As a final note, it can be seen that the phantom-divide curve for
$\gamma < 0$ is a hyperbola, see Eq.~(\ref{geff}); and then
$\gamma_\textrm{eff}$ is positive only well within the $(\Omega_\phi >
1)$-region of the phase plane.

\section{Conclusions}
\label{sec4}
The analysis of the EKG equations from the point of view of dynamical
systems is very fruitful, as the appearance of critical points helps a
lot to reveal the role scalar fields play in many cosmological models.

We have studied the case of a phantom scalar field endowed with an
exponential potential, and found that there are in general only two stable
solutions which are of physical interest. For matter with $\gamma \geq
0$, it is confirmed that the scalar field dominated solution is the
only late-time attractor. 

There is for this case the same problem as in the canonical scalar
field\cite{Copeland:1997et}. In order to have a cosmological evolution
with standard radiation and matter dominated eras, the scalar field
contribution should be almost negligible at early times. If the
phantom scalar field were initially present at the time of inflation, the
inflaton field could not avoid a phantom era unless $\Omega_\phi \ll
1$ at the beginning of inflation. This is because the critical point A
is a saddle point, as we have numerically confirmed, but trajectories
\emph{move} away from it only linearly with $N$.

As in the canonical case, once the radiation and matter dominated eras
appear, the phantom scalar field density parameter grows like
$\Omega_\phi \sim a^{3\gamma}$, which also means that the scalar
energy density remains almost constant until it reaches its late-time
attractor solution. Hence, in order to avoid a disturbing phantom
dominated universe, we need the same constraint that appears in the
standard case, that is, the initial scalar density parameter should be such
that $\Omega_{\phi,i} < \rho_{\gamma ,0}/\rho_{\gamma, i}$, in order
not to become the dominant component at time $t_0$.

The novel case that appears for a phantom equation of state $\gamma$ is the
coexistence of three critical points if $0 > \gamma >
-\lambda^2/3$; a case that can also happen when matters are
coupled\cite{Gumjudpai:2005ry}. Allowing for more than one phantom
type of matter leads to a non-trivial evolution; we learn in this case
that the ultimate phantom regime will depend on the kind of phantoms
present. For instance, it is seen from Fig.~\ref{fig:caseCD} that the
most phantom matter may not be the dominant one at the end. This
result may be relevant in models in which more than one type of phantom
matter coexist together. In\cite{Li:2003ft}, the authors report a
lower bound in the effective equation of state in the case of many
phantom scalars with an ${\cal O}(N)$ symmetry. This suggests that the
classical solutions of phantom scalar fields may have hidden
properties worth investigating.

Another result is that \emph{supra-dominance} of the perfect fluid
$\Omega_\gamma > 1$ cannot be discarded in the case of a phantom
scalar field, because the scalar field contribution $\Omega_\phi$
cannot be, in principle, bounded from below. If phantom scalar fields
were permitted in Nature, there would have to be a mechanism that can
prevent $\Omega_\phi$ to become too negative.

Strangely enough, for $0< \gamma < 2$ the effective equation of state
of the universe $\gamma_\textrm{eff}$ can be negative even in the case
of supra-dominance; moreover, it can change from negative to positive
values, and vice-versa, along some trajectories.

The model presented in this work has the same properties that
appears in any single field-phantom model in which
$\gamma_\phi < 0$: big rip singularity and no crossing of the
$\gamma_\phi=0$ barrier, see\cite{Nojiri:2005sx,Nojiri:2004ip} for
more details.

In studying such properties, the dynamical system formalism
is not only useful but also easy. Apart from helping to decide which
solutions of the EKG system are of physical interest, it also
simplifies the numerical solutions.

The exponential case fits well into the dynamical system formalism, as
has been extensively shown in the literature, but the formalism can be
easily adapted for other scalar potentials. We expect to report on
them elsewhere.

\begin{acknowledgments}
I am thanked to Tonatiuh Matos for sharing his expertise on
Maple, and to an anonymous referee for very helpful suggestions. This
work was partially supported by grants from CONACYT (32138-E, 34407-E,
42748), CONCYTEG (05-16-K117-032), and PROMEP UGTO-CA-3.
\end{acknowledgments}

\bibliography{phantomdsrefs}


\end{document}